\documentclass[12pt]{iopart}  
\usepackage[]{graphicx}
\usepackage{graphicx}
\usepackage{dcolumn}
\usepackage{bm}

\begin{document}

\title{On universality and non-universality for 
a quantum dot in the Kondo regime} 
\author{A. F. Izmaylov$^1$, A. Goker$^2$, B. A. Friedman$^3$ and P. Nordlander$^2$}

\address{$^1$
Department of Chemistry
}
\address{$^2$
Department of Physics and Department of Electrical and Computer Engineering\\
Rice Quantum Institute\\
Rice University\\
Houston, TX 77251-1892, USA
}
\address{$^3$
Department of Physics\\
Sam Houston State University\\
Huntsville, TX 77341, USA
}
\date{\today}

\ead{nordland@rice.edu}

\begin{abstract}
The time-dependent non-crossing approximation is employed for the single-electron transistor
to calculate the transient response of the conductance for a variety of temperatures and biases. 
We consider the case when the dot-lead tunneling constant is suddenly changed such that
the Kondo effect is present in the final state. In the fast non-universal timescale,
where the charge transfer takes place 
, we see rapid oscillations. The frequency of these oscillations
is equal to the dot level and their amplitude is modulated by the initial and final tunneling constants.
To study the slow universal timescale, we develop a new numerical scheme. We 
compute the conductance for two systems which have different 
Kondo temperatures down to a fraction of $T_{K}$ in infinitesimal bias with this scheme. 
We conclude that universality is preserved as a function of $T/ T_{K}$.
We also investigate the decay rate of recently identified split Kondo peak oscillations  
down to zero temperature and compare it
with the previous analytical results obtained with the perturbative renormalization group. 
\end{abstract}


\maketitle

\end{document}